\documentclass[lineno]{jfm}
\usepackage{graphicx}
\usepackage{newtxtext}
\usepackage{newtxmath}
\usepackage{natbib}
\usepackage{hyperref}
\hypersetup{
    colorlinks = true,
    urlcolor   = blue,
    citecolor  = black,
}

\newcommand{\RomanNumeralCaps}[1]
\linenumbers

\title{Regular reflection to Mach reflection (RR-MR) transition in Short wedges}

\author{Vinod Yeldho Baby\aff{1}
  \corresp{\email{grajesh@smail.iitm.ac.in}},
  Vinoth Paramanantham\aff{1}
 \and G. Rajesh\aff{1}}

\affiliation{\aff{1}Indian institute of Technology Madras, Chennai, India.
}

\begin{document}
\maketitle

\begin{abstract}
Regular reflection (RR) to Mach reflection(MR) transitions (RR$\leftrightarrow$MR) on long wedges in steady supersonic flows have been well studied and documented. However, in a short wedge where the wedge length is small, the transition prediction becomes really challenging owing to the interaction of the expansion fan emanating from the trailing edge of the wedge with the incident shock and the triple/reflection point. The extent of this interaction depends on the distance between the wedge trailing edge and the symmetry line (Ht). This distance is a geometric combination of the distance of the wedge leading edge from the symmetry line (H), the wedge angle ($\theta$), and the wedge length (w). In the present work, the RR$\leftrightarrow$MR transitions have been studied on short wedges using analytical and computational methods, and the transition lines are plotted for various Mach numbers and wedge lengths. The transition criterion strongly depends on the wedge length, which can be so adjusted even to eliminate the RR$\leftrightarrow$MR transitions till the wedge angle reaches the no-reflection domain.
\end{abstract}

\begin{figure}
 \centering
 \includegraphics[width=0.72\textwidth]{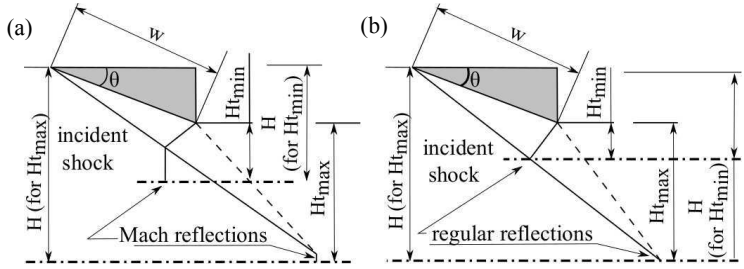}
 \caption{Ht,min and Ht,max for (a) MR and (b) RR.}
 \label{Fig:Ht}
\end{figure}

\section{Introduction}
\label{sec:intro}

\begin{figure}
 \centering
 \begin{minipage}{0.45\textwidth}
 \centering
 \includegraphics[width=.9\textwidth]{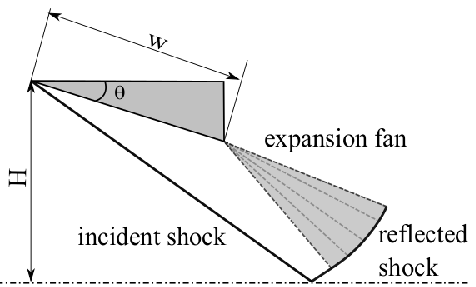}
 \caption{Long wedge.}
 \label{long wedge}
 \end{minipage}
 \begin{minipage}{0.45\textwidth}
 \centering
 \includegraphics[width=0.9\textwidth]{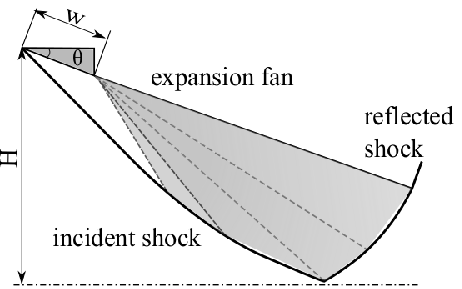}
 \caption{Short wedge.}
 \label{short wedge}
 \end{minipage}
\end{figure}

When a supersonic flow passes over symmetric wedges, a shock interaction/reflection occurs due to the flow turning. The shock reflection can be either a Mach reflection (MR) or a regular reflection (RR), depending on the flow parameters and wedge geometry. Figure \ref{Fig:Ht} depicts the possible shock reflections between the maximum and minimum Ht values, which have been thoroughly investigated \citep{LiBenDor}. The Ht value depends on the wedge length w, wedge angle $(\theta)$ and the centre line distance H of the wedge leading edge, as shown in figures \ref{Fig:Ht} (a) and (b). The shock reflection process is studied thoroughly by varying the wedge angle and flow Mach number. The wedge length is also an essential parameter, and based on the non-dimensionalized wedge length (w/H), we can classify wedges as long or short. As depicted in figure \ref{long wedge}, most research on wedge flow focuses on long wedges where the expansion fan from the trailing edge does not interact with the incident shock wave. As shown in figure \ref{short wedge}, the expansion fan will interact with the incident shock wave in the short wedges. Due to expansion fan interaction, the incident shock angle in short wedges decreases and curves continuously from the point where the leading expansion characteristic meets the shock wave to the triple/reflection point.
The curvature of the incident shock essentially depends on the extent of the expansion fan interaction. Consequently, the incident shock angle at the point of reflection will depend not only on the wedge angle and the flow Mach number but also on the w/H ratio. Any attempt to predict the shock transitions on short wedges must incorporate an accurate model of the incident shock curvature resulting from its interaction with the expansion fan.

\subsection{\label{sec:Lit}Previous works in Shockwave- Expansion fan Interaction}

The interactions of shock waves and expansion waves in one-dimensional flows are the subject of several investigations, in high-speed flows \citep{EMANUEL2001185}. A comprehensive description of the interactions of shock waves, expansion waves and contact surfaces, mainly in shock tube flows, can be seen by the work of \citet{glass1991over}. One of the early works in expansion fan shock wave interaction in symmetric wedge flows was carried out by \citet{LBDor} in which an analytical solution was obtained for the curvature of the shock wave due to its interaction with a centered expansion wave of the opposite family. The curvature of the shock wave was approximated as a second-order polynomial. In the experimental and numerical validations of this analytical model \citep{NelSkews}, it was found that though the model prediction was accurate for small interactions, it could not predict the shock curvature for the stronger interactions. In another study \citep{Hillier} on the shock-wave/expansion-wave interaction and the transition between regular and Mach reflections, analytical and numerical simulations were carried out to investigate the inviscid interactions of an expansion wave with an incident shock wave of the opposite family. The study was to stabilize a Mach reflection in a parallel duct for different flow conditions. Most of the analytical methods developed in shock reflections focused on predicting the MR configuration and the Mach stem height, a finite length scale in an MR in long wedges\citep{azevedo1993engineering,LiBenDor,mouton2007mach}. In these models, the slip line, which originates from the triple point in a Mach reflection, was assumed to be a straight line. Later, the refinement of the slip line curvature was carried out by \citep{Bai2017SizeAS}, resulting in a better Mach stem prediction, as they considered the expansion waves generated over the slip line and its interaction with the transmitted expansion waves through the reflected shock. Further, numerical studies of a shock reflection in the presence of an upstream expansion wave and a downstream shock wave for two-dimensional flows by \citep{yao2013shock} reported that the expansion fan delayed the RR$\leftrightarrow$MR transition at lower Mach numbers (M=2 – 3.5) and expedited it at higher Mach numbers.\\

The analytical models discussed above were all for the interaction of the shock wave with the opposite (either right running or left running) family of expansion waves. On the contrary, in short wedges the incident shock is curved due to the prolonged interaction of the expansion wave of the same family. The study of this type of interaction is significant since the RR$\leftrightarrow$MR transition may solely depend on the incident shock angle at the reflection point, which is determined by the shock curvature due to the interaction. It is hence, essential to determine the parameters governing the shock angle at the reflection point and the change in transition criteria when these parameters vary. No works have been reported till date to investigate the short wedge effects in the RR$\leftrightarrow$MR transitions and the MR configurations, to the best of the authors' knowledge. In the present investigation, an analytical model is hence developed to study the prolonged interaction of an expansion wave with an incident shock wave of the same family to predict the shock wave curvature. This is used to investigate the shock transitions in symmetric short wedges. The analytical model is validated with an in-house inviscid, 2D, structured finite volume solver with a fifth-order WENO scheme \citep{paramanantham_janakiram_gopalapillai_2022}. The RR$\leftrightarrow$MR transition lines are expected to be altered by the interactions of the expansion wave with the shock wave due to the small wedge lengths.

\section{Shock wave expansion fan interaction - Analytical Modeling}\label{MOC}
 The shock–expansion interaction is modeled with the method of characteristics (MoC). The model assumptions are that the flow is steady, compressible, inviscid, two-dimensional (planar), isentropic, and rotational as it is a curved shock. The entropy is constant only along the streamlines. The basic configuration of the incident shock-expansion fan interaction model is shown in figure \ref{MOC}. The characteristic and compatibility equations for two-dimensional, isentropic, rotational, and supersonic flow are given below in Table \ref{tab:eq}.
 
\begin{figure}
 \centering
 \includegraphics{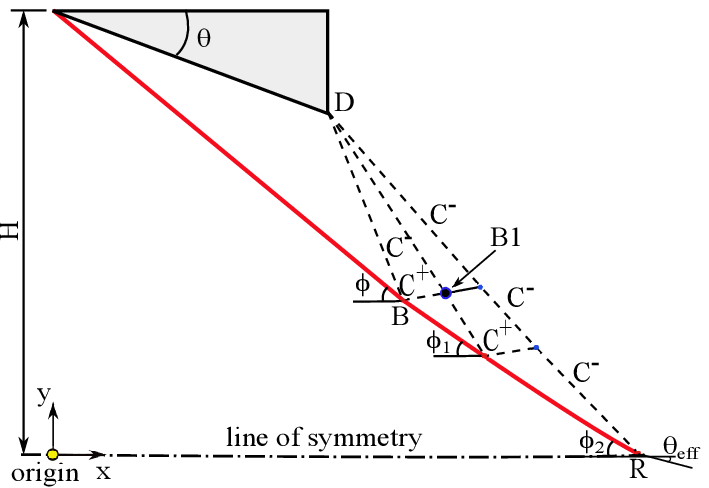}
 \caption{Shock-expansion fan interaction model using MoC.}
 \label{MOCfig}
\end{figure}

 The initial value line for the MOC is the first characteristic expansion line from the trailing edge (point D), which terminates at the intersection of the incident shock wave and the first expansion line (point B). The solution algorithm is initiated by giving a small increment of flow turn angle downstream from point D. The flow properties are determined using the Prandtl-Meyer relations. The C– curve from point D and C+ curve from point B intersect at point B1. The flow properties and location at point B1 are obtained by solving the above-mentioned characteristic and compatibility equations from B and D. The curved shock is obtained by solving discrete points on the shock. Since MoC is not valid across the shock, the solution point at the shock curvature is obtained by iteratively computing the local pressure ratio across the shock. Post-shock properties are calculated with the local incoming Mach number and the assumed pressure ratio across the shock. These flow properties should satisfy the compatibility equation along the C- curve originating from the interior point B1. The iterations stop when this condition is satisfied. The local shock angle is obtained, and the x and y positions of the curved shock are obtained by solving the characteristic equations. Subsequent points on the curved shock are obtained by incrementing the next flow turn angle at point D and repeating the above steps. One of the algorithms calculates the interior points in the flow field, while the other determines the discrete points along the shock curvature. $\phi_2$ and $\theta_{eff}$ represents the shock angle and effective turn angle at the reflection point R respectively as shown in figure \ref{MOCfig}. The analytical model developed does not consider the possibility of Mach reflection and is valid only for regular reflection.
 
 \begin{table}
\begin{center}
\def~{\hphantom{0}}
\begin{tabular}{p{4cm}p{7cm}}

\textbf{Characteristic Equations}
 \\ \hline
 $(dy/dx)   =  \lambda_0  =  v/u$
 &    (Streamline)                     \\
$(dy/dx) = \lambda_+ = tan(\alpha+\mu)$
 & (Right running characteristic line, \textit{‘+’ subscript})                     \\
$(dy/dx) = \lambda_- = tan(\alpha-\mu)$
 & (Left running characteristic line, \textit{‘-’ subscript}) \\  
 \hline
 \textbf{Compatibility Equations}\\ \hline
$\rho VdV+d\rho = 0$ & (Streamline)  \\
$dp - a^2 d\rho = 0$ & (Streamline) \\ 
\\
$\frac{\sqrt{(M^2-1)}}{(\rho V^2 )}  dp_+ +d\alpha_+ = 0$ & (Right running characteristic Mach, \textit{‘+’ subscript}) \\
$\frac{\sqrt{(M^2-1)}}{(\rho V^2 )}  dp_- - d\alpha_- = 0$ & (Left running characteristic line, \textit{‘-’ subscript})
 \end{tabular}
  \caption{Steady,two dimensional, isentropic, rotational supersonic flow}
  \label{tab:eq}
  \end{center}
\end{table}

\begin{figure}
 \centering
  \includegraphics{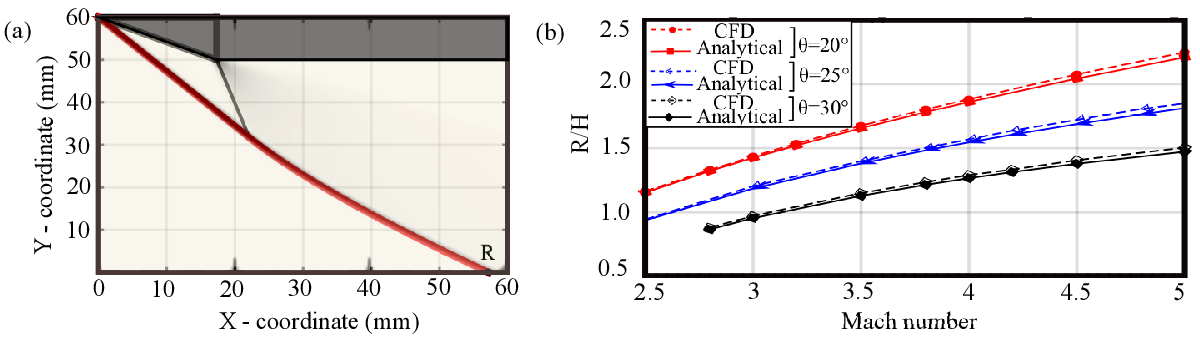}
 \caption{(a) Superimposed image of Analytical and computational result (b) Validation curve for $w/H$=0.33}
 \label{RH-matlab}
\end{figure}

\section {Shockwave expansion fan interaction - Validation}
  The validation of the analytical model with CFD is carried out by superimposing the curved shock locus of the analytical results and the computational results  as shown in figure \ref{RH-matlab}(a). It reveals that the analytical prediction agrees very well with the CFD results qualitatively. The same is done by varying the Mach number, w/H ratio and the wedge angle. The Mach number vs the non-dimensional shock reflection distance [shown in figure \ref{MOC}] (R/H) for different w/H ratios, obtained from analytical and CFD results are plotted. Figure \ref{RH-matlab}(b) shows one such plot of Mach number vs R/H for the w/H ratio 0.33, for different wedge angles. It is seen that the maximum deviation between the analytical and CFD results is less than 2$\%$.

\section{Results and Discussion }\label{Results}

The transition of the shock reflection in a short wedge depends on the extent of the interaction of the expansion fan with the incident shock wave. Due to the interaction of the expansion fan, the incident shock is weakened, resulting in a decrease in the shock angle at the reflection point compared to the case where there is no interaction. Therefore, the transition angle between RR and MR is modified. The value of the shock angle at the point of reflection depends on the degree of interaction of the expansion fan, or the w/H ratio, while M and $\theta$ remain constant. As stated by \cite{LiBenDor}, the transition angle is typically determined using the detachment criterion for RR$\rightarrow$MR or the von Neumann criterion for MR$\rightarrow$RR. In the following sections, we will quantify the transition angles for the short wedges for various w/H ratios using the MoC method described in section 2 and validate these results with higher-order numerical simulations.
\begin{figure}
 \centering
 \includegraphics{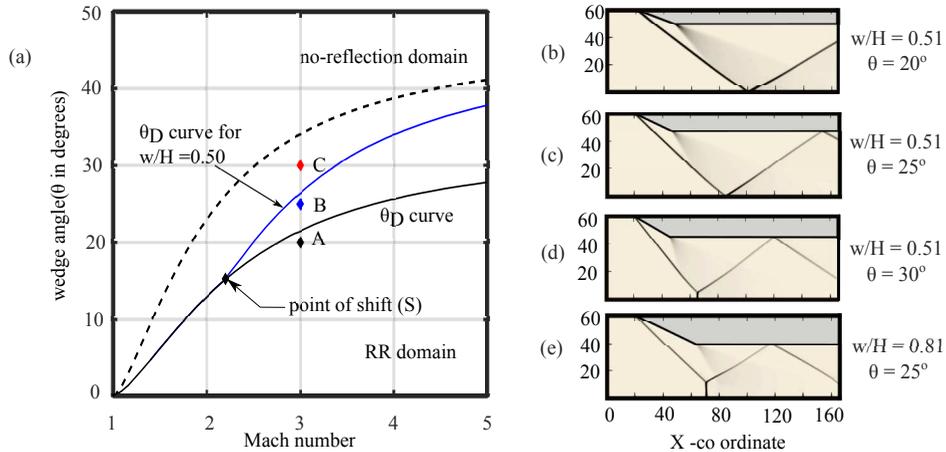}
 \caption{(a) Transition line - Shift in Detachment criteria for w/H = 0.5. (b) Flow configurations corresponding to points A, 
 (c) B, (d) C (all short wedges), and (e) case without interaction(long wedge), where M =3, $w/H = 0.5$ and $\theta = 20 ^{\circ}, 25 ^{\circ}, 30 ^{\circ}$ respectively, for short wedges, and M=3, $w/H = 0.81$ and $25^{\circ}$ for the long wedge.}
 \label{CFD-valid}
\end{figure}

\subsection{Transition lines- Detachment criterion validation}
The delay in the transition criterion, specifically the detachment condition within the short wedge, can be attributed to the attenuation of the incident shockwave. The extent of the delay of this transition is determined by the w/H ratio. The determination of the detachment condition involves the solution of MoC for a specified Mach number (M), wedge angle ($\theta$), and w/H. The detachment condition is generally shifted to higher wedge angles for the shorter wedges and ends at the no-reflection domain. Figure \ref{CFD-valid} (a) shows the shift in the detachment criterion for a w/H ratio of 0.5. The blue line represents the modified detachment criterion resulting from the interaction of the expansion fan with the incident shockwave, whereas the black solid line represents the detachment condition for long wedges in the absence of expansion fan interaction. Point S in figure \ref{CFD-valid} (a) depicts the beginning of the interaction between the expansion fan and the incident shock wave for a given w/H ratio. The deviation of the detachment condition occurs after this value, and for values below the interaction point, it remains the same as the long wedge. The transition line obtained from the MoC is validated with numerical simulations for M = 3 flow over a supersonic wedge; three cases are considered by keeping the Mach number constant and varying the wedge angle to comprehend the type of reflection obtained from the numerical simulations. The wedge angles are set to $\theta_w = 20^{\circ}, 25^{\circ}, 30^{\circ}$  respectively (points A, B, and C in the figure \ref{CFD-valid}(a)), and w/H is set to 0.5 (short wedge case) and 0.81 (long wedge case). Figures \ref{CFD-valid} (b-e) depict the numerical schlierens of the short wedge flow fields at points A, B, and C and the long wedge flow field simulation at point B. It can be seen from figures \ref{CFD-valid} (b-e) that the shock reflections correspond to points A and B are regular reflections,  and C is the Mach reflection for the short wedge, while for point B, it is a Mach reflection for the long wedge. This confirms the shift of the transition line for a short wedge, making a regular reflection at point B, which was an erstwhile Mach reflection for the long wedge. 

\begin{figure}
\centering
\begin{minipage}{0.45\textwidth}
 \centering
 \includegraphics{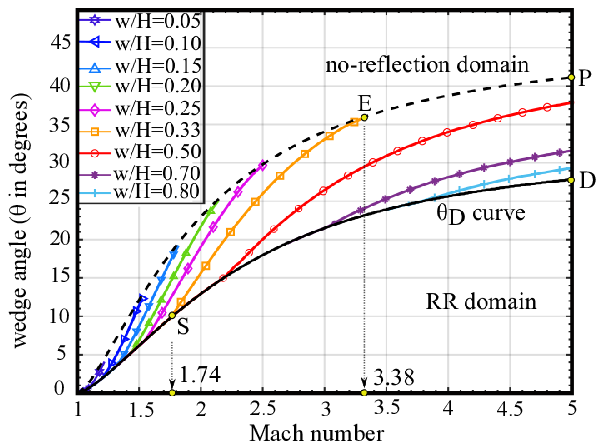}
 \caption{Shift in detachment criterion for various w/H ratios.}
 \label{transline}
 \end{minipage}
  \begin{minipage}{0.4\textwidth}
 \centering
 \includegraphics{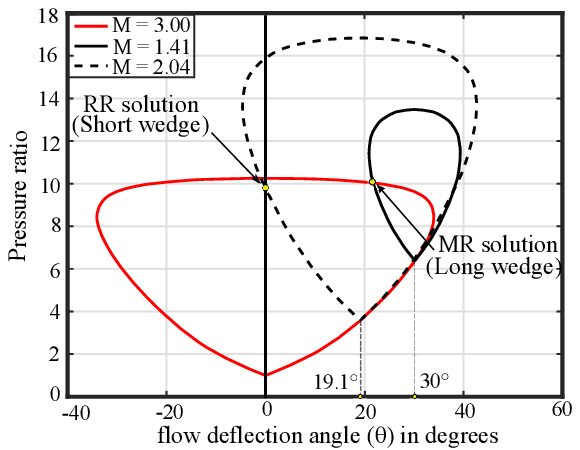}
 \caption{Shock polar representation for w/H = 0.33.}
 \label{shock_polar}
 \end{minipage}
\end{figure}

 Figure \ref{transline} demonstrates that for all w/H ratios, a combination of M-$\theta_w$ exists for which the interaction begins, and the transition line shifts from that point (refer to figure \ref{CFD-valid} (a)). Figure \ref{transline} shows that w/H = 0.33 begins to deviate from the detachment condition line at point S, where  M = 1.74 and $\theta_w = 9.6^{\circ}$. The shift ends at the no-reflection condition at point E corresponding to M = 3.38 and $\theta_w = 36.3^{\circ}$. For w/H = 0.33, the region under the curve SEPD will be RR, which eliminates MR flow configurations for any combination of Mach number and wedge angle beyond point E. Thus, the development of transition conditions for a shorter wedge has far-reaching implications as one can select a domain in which only a regular reflection occurs or control the size of the Mach reflection configuration by adjusting the w/H ratio.

\subsection{Shock polar representation of the interaction effects}

Figure \ref{shock_polar}  depicts the polar shock representation of a long and short wedge with a w/H ratio 0.33. The incoming flow has a Mach number of 3 and a wedge angle of 30$^{\circ}$. At the reflection point, the post-shock Mach number for a long wedge is 1.41, while it is 2.04 for a short wedge. The shock polar yields the MR solution corresponding to the long wedge and the RR solution corresponding to the short wedge. The shock polar reveals that the expansion fan interaction decreases the required flow turn angle at the reflection point in short wedges, even though the wedge angle remains unchanged. 

\subsection{Dual solution domain in short wedges}

\begin{figure}
 \centering
 \includegraphics{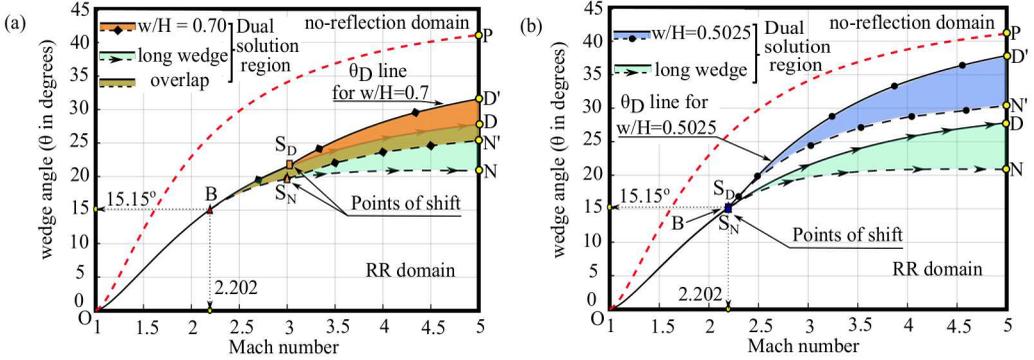}
 \caption{Dual solution domains for short wedges for different w/H ratios (a) w/H=0.7 and long wedge and (b) w/H=0.5025 and long wedge}
 \label{dual_7_5}
\end{figure}

Another significant transition criterion in the wedge flows is the von Neumann criterion. Intuitively, it makes sense to assume that as the w/H ratio decreases, the von Neumann criterion line shifts towards higher wedge angles, similar to the detachment condition. The dual solution domain, the region between the detachment criterion line and the von Neumann criterion line, also undergoes this shift. In order to comprehend the change in the dual solution domain, transition lines for the von Neumann criterion were also drawn for various w/H ratios. The von Neumann criterion line for the short wedge is more complex than the detachment criterion, and the shift in the dual solution domain can be subdivided into three types based on the location of the von Neumann point B, which is the weak Mach reflection point as shown in figure \ref{dual_7_5} (a). We divided the shift in the dual solution domain into three categories as Type I, Type II, and Type III, which are discussed below.
 
\subsubsection{Type I Dual solution domain}

In the type I domain, the von Neumann point 'B' is identical to the long wedge flows, and the von Neumann criterion follows the long wedge von Neumann transition line prior to expansion fan interaction, shifting to higher wedge angles once expansion fan interaction commences. The von Neumann criterion ($\theta_N$) line and detachment criterion ($\theta_D$) line for w/H ratio 0.7 is shown in figure \ref{dual_7_5}(a). The path OBS$_D$D and OBS$_N$N represents the $\theta_D$ and $\theta_N$ lines for long wedges. Both the transition lines follow the same path as that for a long wedge, up to the points of shift (S$_D$ and S$_N$ for the detachment line and von Neumann line, respectively), then follow a different path as (S$_D - $D' and S$_N - $N'). For a short wedge, the path OBS$_D$D' and OBS$_N$N' represents the $\theta_D$ and $\theta_N$ line. The point of the shift of the $\theta_D$ line is at point S$_D$, where M=3.03 and wedge angle $\theta_W =$21.66$^{\circ}$, whereas the point of the shift of the $\theta_N$ line is at point S$_N$, where M=3 and wedge angle $\theta_W =$19.66$^{\circ}$. As the w/H value decreases from 0.7, the point of shifts S$_D$ and S$_N$ moves to the direction of the von Neumann point B, and merges with the point B, where M = 2.202, $\theta_W=$15.15$^{\circ}$ and the w/H ratio is 0.5025 as shown in figure \ref{dual_7_5}(b). Thus, short wedges with w/H ratios greater than 0.5025 will give  a Type I dual solution domain. The characteristic of this type is that the weak Mach reflection or von Neumann point B remains the same for all w/H ratios greater than 0.5025 since both the points of shift from the detachment line and von Neumann line S$_D$ and S$_N$ occur after point B.

\subsubsection{Type II Dual solution domain }

\begin{figure}
 \centering
 \includegraphics{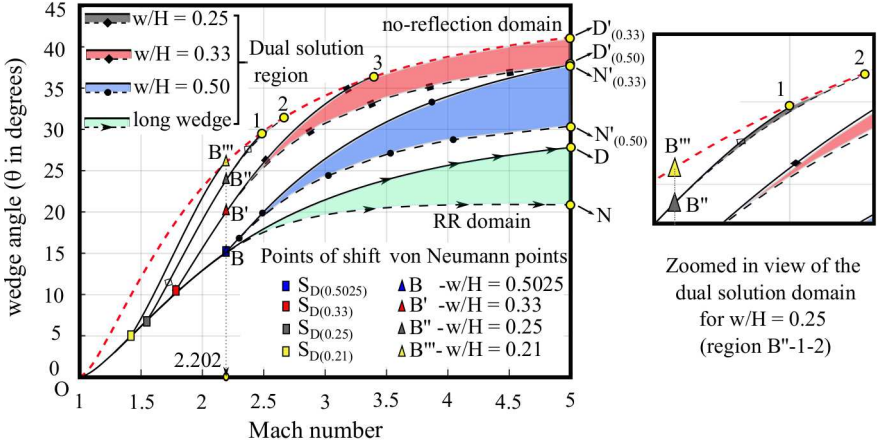}
 \caption{Dual solution domains for long and short wedges}
 \label{dual_sol_all}
\end{figure}

As illustrated in figure \ref{dual_sol_all}, the type II dual solution domain occurs when the detachment condition occurs prior to the von Neumann point B for w/H ratios less than 0.5025.  Consequently, the type II domain begins at w/H = 0.5025 and ends at w/H = 0.21, where the von Neumann point (B) reaches the no-reflection line (point B'''). The dual solution domains for w/H ratios 0.25, 0.33, 0.5025, and long wedge are shown in figure \ref{dual_sol_all}. In the figure, for the long wedges, paths OBD and OBN represent the $\theta_D$ and $\theta_N$ lines, respectively. Path OBD'$_{(0.5)}$, OBN'$_{(0.5)}$, represents the same for short wedge of w/H ratio 0.5025. For w/H ratio 0.33, the $\theta_D$ line and $\theta_N$ line follow the path OB'D'$_{(0.33)}$ and OB'N'$_{(0.33)}$ respectively as shown in figure \ref{dual_sol_all}. The detachment criterion line intersects the no-reflection line at point 3 and then follows the no-reflection line. A major part of the dual solution domain is between the no reflection line and the $\theta_N$ line. The $\theta_D$ line and $\theta_N$ line for w/H ratio 0.25 are shown as path OB''1 and OB''2, respectively. The dual solution domain for w/H=0.25 is extremely small, as seen in the zoomed-in picture in figure \ref{dual_sol_all}. The dual solution domain is hardly visible for wedges with w/H ratios less than 0.25. Thus for short wedges with w/H ratios less than 0.5025, the von Neumann point B shifts in the direction of the no-reflection domain and eventually meets the no-reflection line at point B''', where the $\theta_D$ line of w/H ratio 0.21 intersects with the no-reflection line. Moreover, as the w/H value is reduced from 0.5025 to 0.21, the dual solution domain diminishes and shifts towards the no-reflection domain. 

\subsubsection{Type III Dual solution domain}

The von Neumann solution does not exist for long wedges for  Mach numbers less than 2.202, where the wedge angle $\theta_w=$15.15$^{\circ}$. However, in short wedges, as the w/H ratio decreases, the wedge angle corresponding to the von Neumann point moves to higher wedge angles. The limiting factor to this shift in wedge angle is when von Neumann point intercepts the no-reflection line shown as point B''' in figure \ref{dual_sol_all}. The value of the w/H ratio whose $\theta_D$ line passes exactly through this point B''' will give the limiting point of von Neumann solutions in short wedges. The value calculated  is very close to 0.21, neglecting the numerical errors. The $\theta_D$ line for w/H ratio 0.21 is shown in figure \ref{dual_sol_all} as path OB''', in which point S$_{D(0.21)}$ is the point of the shift from $\theta_D$ line, where M=1.44 and $\theta =$5.3$^{\circ}$. The von Neumann point is at point B''', where M=2.202 and wedge angle $\theta =$25.93$^{\circ}$. Thus the  $M-\theta_w$ parameter space where the value of w/H is less than 0.21 characterizes the type III domain. Type III domain differentiates itself from the other two domains as no dual solution domain exists in this type. 

\subsubsection{Locus of the von Neumann points}

\begin{figure}
 \centering
 \includegraphics{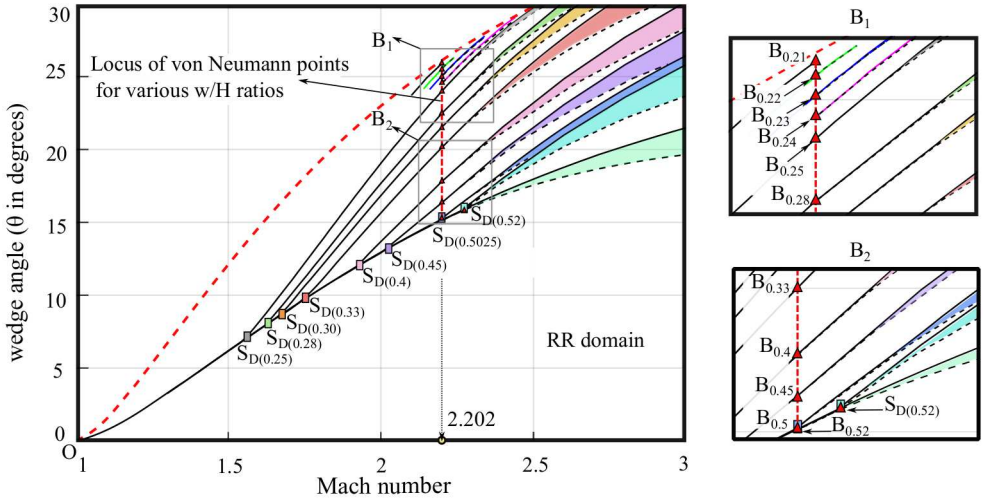}
 \caption{Locus of bifurcation points for various w/H ratio }
 \label{locus of B}
\end{figure}

\begin{figure}
 \centering
 \includegraphics{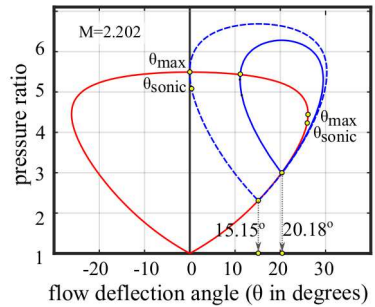}
 \caption{Shock polar for M=2.202, w/H ratio 0.33 }
 \label{shock_polar_locus}
\end{figure}

Shifting the von Neumann point is one of the defining characteristics of the short wedge flows. It is evident from the preceding discussion that for w/H ratios less than 0.5025, the von Neumann point moves, and figure \ref{locus of B} illustrates the locus of shift of the von Neumann points in the red dashed line for various w/H ratios. The red dashed line divides the region between the detachment and no-reflection lines into two distinct areas. Dual-domain solutions do not exist on the left side of the locus of the line, but they do exist on the right. The shift in von Neumann points in the $M-\theta_w$ parameter space follows a straight line passing through the X-intercept where M=2.202. Although the dual solution domain exists on the right side of the above-mentioned straight line, the domain close to this line is hardly visible due to the interaction effects, as shown in figure \ref{locus of B}. To confirm the shift in von Neumann point B, shock polars of short wedges with actual wedge angle and effective wedge angle(turn angle at reflection point) are drawn. Figure \ref{shock_polar_locus} shows the shock polar for w/H ratio 0.33 at the von Neumann point B$_{0.33}$ (refer figure \ref{locus of B}). The solid blue line indicates r-polar for the actual wedge angle used, and the dashed blue line shows the same for the effective wedge angle $\theta_{eff}$ at the reflection point R(refer figure \ref{MOC}). At point B$_{0.33}$, the wedge angle is 20.18$^{\circ}$ and due to interaction this is reduced to 15.15$^{\circ}$. Thus the von Neumann shift in short wedge flows is validated with the shock polars. Further investigations are required to comprehend the shock reflection configurations and the flow behaviour in Type I, Type II and Type III interactions in the flow over short wedges.

\section{Conclusion}
The shock reflection phenomena on short wedges have been investigated analytically and computationally. The interaction of the expansion fan emanating from the trailing edge of the short wedge with the incident shock wave is found to curve the incident shock. The curving of the incident shock wave leads to the change of shock angle at the reflection/triple point, consequently leading to a shift in the transition line.  The transition lines plotted for different w/H ratios from 0.05 to 0.8 show that the shift in transition lines depends on the extent of interaction. The plot is useful in finding the detachment/von Neumann criterion for any Mach number between 1 and 5 for a particular w/H ratio. Moreover, we can control the appearance of MR or its configuration, for a particular Mach number by selecting the suitable (w/H) ratio from the transition lines generated. As the extent of expansion fan interaction with the incident shock increases, the dual domain shifts its location and diminishes as a function of w/H ratio facilitating the control of the MR$\leftrightarrow$RR hysteresis process.
\bibliographystyle{jfm}
\bibliography{jfm}

\begin{thebibliography}{11}
\expandafter\ifx\csname natexlab\endcsname\relax\def\natexlab#1{#1}\fi
\def\au#1{#1} \def\ed#1{#1} \def\yr#1{#1}\def\at#1{#1}\def\jt#1{\textit{#1}}
  \def\bt#1{#1}\def\bvol#1{\textbf{#1}} \def\vol#1{#1} \def\pg#1{#1}
  \def\publ#1{#1}\def\arxiv#1{#1}\def\org#1{#1}\def\st#1{\textit{#1}}

\bibitem[Azevedo \& Liu(1993)]{azevedo1993engineering}
{\sc \au{Azevedo, D.~J.} \& \au{Liu, Ching~Shi}} \yr{1993}  \at{Engineering
  approach to the prediction of shock patterns in bounded high-speed flows}.
  \jt{AIAA Journal}  \bvol{31}~(1),  \pg{83--90},  \arxiv{arXiv:
  https://doi.org/10.2514/3.11322}.

\bibitem[Bai \& Wu(2017)]{Bai2017SizeAS}
{\sc \au{Bai, Chen-Yuan} \& \au{Wu, Zi-Niu}} \yr{2017}  \at{Size and shape of
  shock waves and slipline for mach reflection in steady flow}.  \jt{Journal of
  Fluid Mechanics}  \bvol{818},  \pg{116–140}.

\bibitem[Emanuel(2001)]{EMANUEL2001185}
{\sc \au{Emanuel, George}} \yr{2001}  \at{Chapter 3.1 - shock waves in gases}.
  \bt{In {\em Handbook of Shock Waves, Vol-1\/} (ed. \ed{G~BEN-DOR, O~IGRA \&
  T~ELPERIN})},  \pg{pp. 185--262}.  \publ{Burlington: Academic Press}.

\bibitem[Glass(1991)]{glass1991over}
{\sc \au{Glass, I.~I.}} \yr{1991}  \at{Over forty years of continuous research
  at utias on nonstationary flows and shock waves}.  \jt{Shock Waves}
  \bvol{1}~(1),  \pg{75--86}.

\bibitem[Hillier(2007)]{Hillier}
{\sc \au{Hillier, R.}} \yr{2007}  \at{Shock-wave/expansion-wave interactions
  and the transition between regular and mach reflection}.  \jt{Journal of
  Fluid Mechanics}  \bvol{575},  \pg{399–424}.

\bibitem[Li \& Ben-Dor(1996)]{LBDor}
{\sc \au{Li, H.} \& \au{Ben-Dor, G.}} \yr{1996}
  \at{Oblique-shock/expansion-fan interaction - analytical solution}.  \jt{AIAA
  Journal}  \bvol{34}~(2),  \pg{418--421},  \arxiv{arXiv:
  https://doi.org/10.2514/3.13081}.

\bibitem[Li \& Ben-Dor(1997)]{LiBenDor}
{\sc \au{Li, H.} \& \au{Ben-Dor, G.}} \yr{1997}  \at{A parametric study of mach
  reflection in steady flows}.  \jt{Journal of Fluid Mechanics}  \bvol{341},
  \pg{101–125}.

\bibitem[Mouton \& Hornung(2007)]{mouton2007mach}
{\sc \au{Mouton, Christopher~A.} \& \au{Hornung, Hans~G.}} \yr{2007}  \at{Mach
  stem height and growth rate predictions}.  \jt{AIAA Journal}  \bvol{45}~(8),
  \pg{1977--1987},  \arxiv{arXiv: https://doi.org/10.2514/1.27460}.

\bibitem[Nel {\em et~al.\/}(2015)Nel, Skews \& Naidoo]{NelSkews}
{\sc \au{Nel, Lara}, \au{Skews, Beric} \& \au{Naidoo, Kavendra}} \yr{2015}
  \at{Schlieren techniques for the visualization of an expansion fan/shock wave
  interaction}.  \jt{Journal of Visualization}  \bvol{18}~(3),  \pg{469--479}.

\bibitem[Paramanantham {\em et~al.\/}(2022)Paramanantham, Janakiram \&
  Gopalapillai]{paramanantham_janakiram_gopalapillai_2022}
{\sc \au{Paramanantham, Vinoth}, \au{Janakiram, Sushmitha} \& \au{Gopalapillai,
  Rajesh}} \yr{2022}  \at{Prediction of mach stem height in compressible open
  jets. part 1. overexpanded jets}.  \jt{Journal of Fluid Mechanics}
  \bvol{942},  \pg{A48}.

\bibitem[Yao {\em et~al.\/}(2013)Yao, Li \& Wu]{yao2013shock}
{\sc \au{Yao, Y.}, \au{Li, S.~G.} \& \au{Wu, Z.~N.}} \yr{2013}  \at{Shock
  reflection in the presence of an upstream expansion wave and a downstream
  shock wave}.  \jt{Journal of Fluid Mechanics}  \bvol{735},  \pg{61–90}.

\end{thebibliography}

\end{document}